\documentstyle[amstex,12pt,twoside]{article}            %% AMS-LaTeX
\newlength{\dinwidth}
\newlength{\dinmargin}
\newtheorem{Definition}{Definition}[section]

\newtheorem{Proposition}[Definition]{Proposition}
\newtheorem{Lemma}[Definition]{Lemma}

\newsymbol\rest 1316         %% restriction symbol
\newcommand{\RR}{{\Bbb R}}                      %real numbers
\begin{document}
\title{On the Implementation of Supersymmetry\\[2mm]
{\small Dedicated to Jan {\L}opuszanski}}
\author{Detlev Buchholz
\\[2mm]
Institut f\"ur Theoretische Physik, Universit\"at 
G\"ottingen,\\ Bunsenstra{\ss}e 9, D-37073 G\"ottingen, Germany\\[0mm]} 
\date{}
\maketitle
\begin{abstract}{\noindent}The implementation 
of supersymmetry transformations by Hilbert space operators 
is discussed in the framework of supersymmetric C$^*$--dynamical 
systems. It is shown that the only states admitting such an implementation 
are pure supersymmetric ground states or 
mixtures and elementary excitations thereof. Faithful states,
such as KMS--states, are never supersymmetric. 
\end{abstract}

\section{Introduction}
\setcounter{equation}{0}
Supersymmetry is an intriguing mathematical concept 
which has become a basic ingredient in many branches of 
modern theoretical physics. In spite of its still lacking 
physical evidence, its far--reaching theoretical implications
uphold the believe that supersymmetry plays a prominent 
role in the fundamental laws of nature. 

As for the theory of elementary particles, the possible manifestations
of unbroken supersymmetry have been fully clarified by 
Haag, {\L}opuszanski and Sohnius \cite{HaLoSo,Lo}. On the other hand 
it is known that supersymmetry is inevitably broken in thermal 
states. As a matter of fact, this breakdown is much 
stronger than that of internal bosonic symmetries: one may speak 
of a spontaneous collapse of supersymmetry \cite{BuOj}. 

These facts seem to indicate that supersymmetry is only 
implementable in states describing elementary systems. 
It is the aim of the present article to clarify this point 
for general C$^*$--dynamical systems.
Apart from supersymmetry, the only ingredient in our analysis is 
the assumption that the dynamics is asymptotically abelian
(see below for precise definitions). So our framework 
covers also non--local theories. 

We shall show in the subsequent section that supersymmetric states 
are always ground states. If these states are mixed (not pure),  
they can be decomposed into pure 
states which are also supersymmetric. At the other extreme, 
faithful states (such as KMS--states) are never supersymmetric.
States which are not supersymmetric but still admit an implementation 
of the supersymmetry transformations 
by Hilbert space operators coincide asymptotically with supersymmetric
ground states and may thus be regarded as excitations
thereof. The physical significance of these 
results is discussed in the conclusions. 
\section{Implementations of odd derivations}
\setcounter{equation}{0}
We discuss in this section the consequences of unbroken 
supersymmetry. As our results 
do not rely on a specific physical interpretation we present them 
in the general mathematical setting of 
C$^*$--dynamical systems \cite{Sa}.\\[2mm]  
{\bf Definition:} Let ${\cal F} = {\cal F}_+ \oplus {\cal F}_-$ be a 
graded C$^{\, *}$--algebra, let $\alpha_t, \, t \in \RR,$ be a 
group of automorphisms of $\cal F$ which respects the grading
and let ${\cal A} \subset {\cal F}$ be the dense subalgebra of 
analytic elements with respect to the action of $\alpha$.   
The dynamical 
system $({\cal F}, \alpha)$ is said to be supersymmetric if the 
(skew symmetric) generator of $\alpha$
\begin{equation}
\delta_0 \doteq -i 
\mbox{\large $\frac{d}{dt}$} \alpha_t \ |_{t=0} \label{2.1}  
\end{equation} 
can be represented in the form 
\begin{equation}
\delta_0 = \mbox{\large $\frac{1}{2}$}
(\delta \cdot \overline{\delta} + \overline{\delta} \cdot \delta ), \label{2.2}
\end{equation} 
where $\delta$ is a closable odd derivation which is defined on 
${\cal A}$ and commutes with $\alpha$, and the associated linear map  
$\overline{\delta}$ on ${\cal A}$ is fixed by 
\begin{equation}
\overline{\delta} (F_\pm) \doteq \mp \, \, \delta ({F_\pm}^*)^* 
\label{2.3}
\end{equation} 
for $F_\pm \in {\cal A}_\pm \doteq {\cal A} \cap {\cal F}_\pm$. 

The even and odd parts ${\cal F}_\pm$ of $\cal F$ 
may be interpreted as the Bose
and Fermi parts of some field algebra. There holds
in particular ${\cal F}_+ \cdot {\cal F}_- = {\cal F}_- \cdot {\cal
  F}_+ \subset {\cal F}_-$ and ${\cal F}_\pm \cdot {\cal F}_\pm 
\subset {\cal F}_+$. We recall that  
an odd derivation is a densely defined linear mapping which 
maps even operators into odd ones and {\em vice versa}, and which 
satisfies the graded Leibniz rule 
\begin{equation}
\delta (F_\pm \, G) = \delta (F_\pm) \, G 
\pm \ F_\pm \, \delta (G) \label{2.4}
\end{equation} 
for $F_\pm \in {\cal A}_{\pm}$ and $G \in {\cal A}$. It is easily
checked that $\overline{\delta}$ is also an odd derivation.

Note that the right hand side of relation (\ref{2.2}) 
always defines an even derivation.  
Hence, given $\delta$, one can determine a corresponding 
$\delta_0$ and if the latter derivation is sufficiently well behaved 
it is the generator of a 
group of automorphisms $\alpha$ satisfying relation 
(\ref{2.1}) \cite[Ch.\ 3.4]{Sa}. In this sense  
the whole structure is fixed by $\delta$. 
We turn now to the analysis of supersymmetric states.\\[3mm]
{\bf Definition:} A state $\omega$ on $\cal F$ is said to be
supersymmetric if $\omega \! \cdot \! \delta = 0$.\\[-2mm]

The following result on the implementability of derivations 
in representations induced by symmetric states  
is well known in the even case \cite{Sa}. Its 
straightforward generalization to odd derivations 
is given here for completeness.
\begin{Lemma}  
Let $\omega$ be a supersymmetric state on $\cal F$ and 
let  $(\pi,{\cal H},\Omega)$ be its  
induced GNS--representation.
The operator $Q$ given by  
\begin{equation} 
Q \, \pi (F) \, \Omega \doteq \pi (\delta (F)) \, \Omega \quad \mbox{for} 
\quad F \in {\cal A}\label{2.5}
\end{equation}
is well defined and closable. Moreover, there holds on its domain 
$\pi ({\cal A}) \, \Omega $  
\begin{equation} 
\pi (\delta (F_\pm)) = Q \pi (F_\pm ) \mp \pi (F_\pm) Q \quad \mbox{for} 
\quad F_\pm \in {\cal A}_{\pm}. \label{2.6}
\end{equation}
\end{Lemma}
{\em Proof:\/} By relation
(\ref{2.4}) and the supersymmetry of $\omega$
there holds for any $F \in {\cal A}$ 
and $G_\pm \in {\cal A}_{\pm}$
\begin{eqnarray} 
(\pi (G_\pm) \, \Omega, \, \pi (\delta (F)) \, \Omega) & = &
\omega ({G_\pm}^* \, \delta (F)) = 
\omega (\mp \, \delta ({G_\pm}^*) \, F) \nonumber \\ & = &  
(\pi(\mp\,\delta({G_\pm}^*)^*) \, \Omega, \, \pi (F) \, \Omega)
\nonumber \\ & = &
(\pi(\overline{\delta}({G_\pm})) \, \Omega, \, \pi (F) \, \Omega). \label{2.7} 
\end{eqnarray}
As in the case of even derivations one concludes from this equality that 
$Q$ is a well defined linear operator which is closable. In fact, its
adjoint $Q^{\, *}$ is also defined on $\pi ({\cal A}) \, \Omega $  
and $Q^{\, *} \, \pi (F) \, \Omega = \pi(\overline{\delta}(F)) \, \Omega$, 
$F \in {\cal A}$. The second part of the statement follows from 
relations (\ref{2.4}) and (\ref{2.5}) 
after a routine computation. \hfill $\Box$\\[-2mm]

Next we show, by making use of arguments in \cite{BuOj}, that  
supersymmetric states are ground states with respect to the 
group $\alpha$. If they are mixed, 
all states appearing
in their decomposition are also supersymmetric. 
\begin{Proposition} Let $\omega$ be a supersymmetric 
state on $\cal F$. Then the group of automorphisms 
$\alpha$ is implemented in the corresponding GNS--representation 
$(\pi,{\cal H},\Omega)$ 
by a continuous unitary group $U$ with positive generator and 
$\Omega$ is invariant under the action of $U$. If 
$\omega$ is a mixed  state, any component (sub--ensemble)
$\omega_{\scriptscriptstyle <}$ 
appearing in its decomposition is also supersymmetric. 
\end{Proposition}
{\em Proof:\/} As $\omega \cdot \delta = 0$ there holds 
$\omega \cdot \overline{\delta}  = 0 $, hence $\omega \cdot
\delta_0 = 0 $. It therefore follows from standard
arguments that $\alpha_t, \, t \in \RR$, is implemented by a continuous 
unitary group $U(t), \, t \in \RR$, which leaves $\Omega$ invariant. Now 
for any $\sigma, \tau \in \{ \pm \}$ and  
$F_\sigma \in {\cal A}_\sigma, \, G_\tau \in {\cal A}_\tau$ we have 
\begin{eqnarray}
{\delta} ({F_\sigma}^* \, \overline{\delta} (G_\tau)     ) & = & 
{\delta} \, ({F_\sigma}^*) \, \overline{\delta} (G_\tau) + 
\sigma \, {F_\sigma}^* \, 
    ( \delta \cdot \overline{\delta} \, (G_\tau)     ) \nonumber \\
& = & - \sigma \, \overline{\delta} \, (F_\sigma)^* \, 
\overline{\delta} (G_\tau) + \sigma \, {F_\sigma}^* \, 
     ( \delta \cdot \overline{\delta} \, (G_\tau)     ),
\label{2.8} 
\end{eqnarray}
and consequently 
$\omega ({F_\sigma}^* \, 
( \delta \cdot \overline{\delta} \, (G_\tau) )) = 
\omega (\overline{\delta} (F_\sigma){}^* \, \overline{\delta}
(G_\tau))$. 
By interchanging the role of $\delta$ and $\overline{\delta}$ we also get 
$\omega ({F_\sigma}^* \, 
( \overline{\delta}  \cdot {\delta} \, (G_\tau) )) = 
\omega ({\delta} (F_\sigma){}^* \, {\delta}
(G_\tau))$. 
Since $\delta, \overline{\delta}$ are 
linear we conclude that for $F \in {\cal A}$
\begin{eqnarray}
\omega ({F}^*  {\delta}_0 (F)) & = & 
\mbox{\large $\frac{1}{2}$} \,  
\omega ({F}^* 
( \delta \cdot \overline{\delta} \, (F) )) + 
\mbox{\large $\frac{1}{2}$} \, 
\omega ({F}^*  
( \overline{\delta}  \cdot {\delta} \, (F) ))
\nonumber \\ & = & 
\mbox{\large $\frac{1}{2}$} \, \omega (\overline{\delta} (F)^* \, 
\overline{\delta}  (F)) + 
\mbox{\large $\frac{1}{2}$} \, \omega ({\delta} (F)^* \, {\delta} (F))
\geq 0, \label{2.9}
\end{eqnarray}
proving that the generator of $U$ is a positive selfadjoint operator
\cite[Ch.\ 4.2]{Sa}.

Next, if $t \mapsto f(t)$ is any absolutely integrable function whose 
Fourier transform has support in $\RR_-$ and if $F \in {\cal F}$ we
put $\alpha_f (F) \doteq \int  \! dt \, f(t) \, \alpha_t (F)$. 
Since $t \mapsto \alpha_t (F)$ is strongly continuous there holds 
$\alpha_f (F)\in 
{\cal F}$. It follows from the preceding result that 
$\omega (\alpha_f (F)^* \, \alpha_f (F)) = 0$. Hence  
if $\omega_{\scriptscriptstyle <} \leq c \cdot
\omega$ for some positive constant $c$, there holds  
$\omega_{\scriptscriptstyle <} (\alpha_f (F)^* \, \alpha_f (F)) = 0$
and consequently $\omega_{\scriptscriptstyle <}$ is also invariant under 
the action of $\alpha$. Setting $F_T \doteq 
\mbox{$T^{-1}$} 
\int_0^T \! dt \, \alpha_t (F), \, F \in {\cal A}$, we get
$\omega_{\scriptscriptstyle <} (\delta (F)) = 
\omega_{\scriptscriptstyle <} (\delta (F)_T) =
\omega_{\scriptscriptstyle <} (\delta (F_T))$
and making use of relation (\ref{2.9})
we obtain the inequality
\begin{eqnarray}
|\omega_{\scriptscriptstyle <} (\delta (F))|^2 & = & 
|\omega_{\scriptscriptstyle <} (\delta (F_T))|^2 \leq
\omega_{\scriptscriptstyle <} (\delta (F_T)^* \, \delta (F_T) ) 
\nonumber \\ 
& \leq & c \cdot \omega  (\delta (F_T)^* \, \delta (F_T) ) +
 c \cdot \omega  ( \overline{\delta} (F_T)^* \, \overline{\delta} (F_T) )
\nonumber \\
& = & 2 c \cdot \omega  (F_T^* \, \delta_0 (F_T) ). \label{2.10}
\end{eqnarray} 
As $\delta_0 (F_T) = -i \mbox{$T^{-1}$} (\alpha_T (F) - F)$,
the right hand side of this inequality tends to $0$ as $T \rightarrow 
\infty$ and the assertion follows. \hfill $\Box$ 
\vspace*{2mm}

The following proposition is a straightforward consequence of this result.
\begin{Proposition} If $\omega$ is a faithful state on $\cal F$
and $\delta \neq 0$, there holds $\omega \cdot \delta \neq 0$. 
\end{Proposition}
{\em Proof:\/} If $\omega \cdot \delta = 0$,  
it follows from the preceding proposition that 
for any absolutely integrable function $f$ whose 
Fourier transform has support in $\RR_-$ and any 
$F \in {\cal F}$ there holds 
$\omega (\alpha_f (F)^* \, \alpha^{}_f (F)) = 0$. 
Since $\omega$ is faithful this implies 
$\alpha_f (F) = 0$ and 
$\alpha_{\, \overline{f}} (F^*) = \alpha_{f} (F)^* = 0$. As   
the Fourier transform of the complex conjugate $\overline{f}$ of $f$
has support in $\RR_+$ and 
$f, F$ are arbitrary we arrive at $\alpha_f (F) = 0$ 
whenever the Fourier transform of $f$ does not contain $0$ in its 
support. Hence
$\alpha_t (F) = F$ for $ t \in \RR$ and consequently $\delta_0 = 0$.  
Because of relation (\ref{2.9}) this implies 
$\omega (\delta(F)^* \delta(F)) = 0$ 
for all $F \in {\cal A}$. But this 
is incompatible with the assumption that $\omega$ is faithful and  
$\delta \neq 0$. Hence $\omega \cdot \delta \neq 0$. \hfill $\Box$ 
\vspace*{2mm}

More can be said if there acts on $\cal F$ 
some group of automorphisms in an asymptotically abelian manner.
In order to simplify the discussion we assume that $\alpha$ 
itself has this property and indicate below
which of the subsequent results hold more generally.\\[2mm]
{\bf Definition:} The group $\alpha_t, \, t \in \RR$, 
is said to act on $\cal F$ in an asymptotically abelian manner 
(shortly: it is asymptotically abelian) if 
\begin{equation}
|| \alpha_t (F_\pm) \, G_\pm \mp \, G_\pm \, \alpha_t (F_\pm) 
|| \rightarrow 0 \label{2.11}
\end{equation}
for $F_\pm, G_\pm \in {\cal F_\pm}$ and $|t| \rightarrow \infty$.\\[-2mm]

A well known consequence of asymptotic
abelianess is the following result on the asymptotic
behaviour of averages of odd operators \cite{BuOj,NaTh}. 
\begin{Lemma} Let $\alpha_t, \, t \in \RR,$ be asymptotically abelian. Then 
\begin{equation}
\lim_{T \rightarrow \infty} \, || \mbox{$T^{-1}$}  
\int_0^T \! dt \, \alpha_t (F_-) || = 0 \quad \mbox{for} \quad 
F_- \in {\cal F}_-. \label{2.12}
\end{equation} 
\end{Lemma}
{\em Proof:\/} Since for $F \in {\cal F}$ there holds 
$||F||^2 = ||F^* F|| \leq ||F^* F + F F^* ||$ we obtain the estimate
\begin{eqnarray}
|| \mbox{$T^{-1}$} \! \int_0^T \! dt \, \alpha_t (F_-) ||^2 & \leq & 
|| \mbox{$T^{-2}$} \! \int_0^T \! dt  \int_0^T \! dt^\prime 
\,     (\alpha_t (F_-)^*  \alpha_{t^\prime} (F_-) +
\alpha_{t^\prime} (F_-) \alpha_t (F_-)^*     )||  \nonumber \\
& \leq & \mbox{$T^{-2}$}  \! \int_0^T \! dt \int_{-T}^T \! dt^\prime \, \,
|| {F_-}^*  \, \alpha_{t^\prime} (F_-) + \alpha_{t^\prime} 
(F_-) \, {F_-}^* ||.   \label{2.13}
\end{eqnarray} 
According to relation (\ref{2.11}) the norm under the integral on the 
right hand side of this inequality tends to $0$ if $|t^\prime|
\rightarrow \infty$, so the statement follows. \hfill $\Box$\\[-2mm]

With these preparations we can establish now more detailed 
information on the representations induced by 
supersymmetric states. Our result relies on  
familiar arguments in algebraic quantum field theory \cite{Sa}. 
\begin{Proposition}
Let the group $\alpha_t, t \in \RR$, be asymptotically
abelian. If the state $\omega$ is supersymmetric,    
the GNS--representation $\pi$ of $\cal F$ induced by
$\omega$ is of type I. More specifically, the commutant 
$\pi ({\cal F})^{\prime}$ of $\pi ({\cal F})$ coincides with  
the center of $\pi ({\cal F})^{\prime \prime}$. 
\end{Proposition}
{\em Proof:\/} According to Proposition 2.2 the automorphisms  
$\alpha$ are implemented in the GNS--repre\-sen\-tation 
$(\pi,{\cal H},\Omega)$ 
by a continuous unitary group $U$ with positive generator, and
$\Omega$ is invariant under the action of $U$. Hence 
$U(t) \in \pi ({\cal F})^{\prime \prime}$ for $t \in \RR$  
by a theorem of Araki \cite[Ch.\ 2.4]{Sa}. Let $E_0$ be the 
projection onto the $U$--invariant subspace in $\cal H$. 
The statement follows if 
$E_0 \pi ({\cal F}) E_0$ can be shown to be 
a commutative family \cite[Prop.\ 2.4.11]{Sa}. Now by the preceding
Lemma there holds for $F_- \in {\cal F}_-$
\begin{equation}
E_0 \, \pi({F}_-) E_0 = 
 E_0  \, \pi (\mbox{
$T^{-1}$} \! \int_0^T \! dt \, \alpha_t (F_-)) E_0 
\rightarrow 0 \label{2.14}
\end{equation}
as $T \rightarrow \infty$ and consequently 
$E_0 \, \pi ({\cal F}_-) E_0 = 0$. Similarly, if 
$F_+, G_+ \in {\cal F}_+$, we obtain from relation 
(\ref{2.11}) by standard arguments (mean ergodic theorem)
\begin{equation}
[ E_0 \, \pi({F}_+) E_0,  E_0 \, \pi({G}_+) E_0 ]
= \lim_{T \rightarrow \infty} \! 
\mbox{ $T^{-1}$} \! \int_0^T \! dt \, E_0 \, 
\pi     ( [ {F}_+, \alpha_t (G_+) ]     ) E_0 = 0, \label{2.15}
\end{equation}
where the convergence is understood in the weak operator topology. \hfill 
$\Box$\\[-2mm] 

The preceding result and the second part of Proposition 2.2 imply that any 
supersymmetric state is a pure supersymmetric ground state or a 
mixture of such states.
Hence supersymmetric states describe the most
elementary systems of the theory. Next we analyze the class of 
states which are not supersymmetric but still admit an implementation of 
supersymmetry transformation by Hilbert space operators.\\[2mm]
{ \bf Definition:}
A state $\omega$ on $\cal F$ is said to be super--regular if  
$\delta$ is implementable in its 
GNS--representation $(\pi,{\cal H},\Omega)$, i.e., if there 
is some densely defined, closed operator $Q$ on $\cal H$ such that 
there holds in the sense of sesquilinear forms 
\begin{equation}
\pi ( \delta (F_\pm)) = Q \pi (F_\pm) \mp \pi (F_\pm) Q
\quad \mbox{for} \quad F_\pm \in {\cal A}_\pm. \label{2.16}
\end{equation} 
Note that it is not required that $\Omega$ is contained in the 
domain of $Q$. \\[-2mm]

In order to proceed we need the following technical result 
which seems of interest in its own right. In its proof 
we apply similar arguments as in the analysis of even 
derivations in \cite{BuDoLoRo}.  
\begin{Lemma}
If $\omega$ is a super--regular state on $\cal F$,  
there is for any $\varepsilon > 0$ a constant 
$c_\varepsilon$ such that for all $F_\pm \in {\cal A}_\pm$ 
\begin{equation}
|\, \omega (\delta (F_\pm) ) | \leq c_\varepsilon 
    ( || \pi (F_\pm) \Omega || + || \pi ({F_\pm}^*) \Omega||     ) 
+ \varepsilon       ( || \delta (F_\pm)|| + ||\overline{\delta}
(F_\pm)||). \label{2.17}
\end{equation} 
\end{Lemma}
{\em Proof:\/} We begin by recalling that if  $Q$ is a  
densely defined, closed operator its
adjoint $Q^*$ has the same property and the operators $Q Q^*$ and 
$Q^* Q$ are selfadjoint and positive. {}From the equality 
\begin{equation}
(\Phi, \pi(\overline{\delta} (F_\pm)) \Psi)
= \mp \, \overline{(\Psi, \pi({\delta} ({F_\pm}^*) \Phi))} \, , \label{2.18} 
\end{equation} 
where $F_\pm \in {\cal A}_\pm$ and $\Phi, \Psi$ are vectors in the 
domains of $Q$ and $Q^*$, respectively, 
it follows that $\overline{\delta}$
is also implementable and 
\begin{equation}
\overline{\delta} (F_\pm) = Q^* F_\pm \mp  F_\pm Q^*  \label{2.19}
\end{equation} 
in the sense of sesquilinear forms. Putting for $\eta > 0$
\begin{equation}
L_\eta \doteq (1 + i \eta \, Q Q^*)^{-1}, \quad 
R_\eta \doteq (1 + i \eta \, Q^* Q)^{-1} \label{2.20}
\end{equation} 
it is clear that $L_\eta$, $R_\eta$ as well as 
$L_\eta Q \subset (Q^* L_\eta^*)^*$ 
and $Q R_\eta$ are bounded operators. Moreover, 
by making use of relations (\ref{2.16}) and (\ref{2.19}) one finds  
after a straightforward computation that for $F_\pm \in {\cal A}_\pm$   
\begin{equation}
 L_\eta Q \, \pi(F_\pm) \mp \pi(F_\pm) \, Q R_\eta =
L_\eta  \, \pi (\delta (F_\pm))  \, R_\eta 
\mp i \eta  \, L_\eta Q  \, \pi ( \overline{\delta} (F_\pm))  \, Q  R_\eta.
\label{2.21}
\end{equation} 
Now with the help of the spectral theorem one sees that the 
norms of the vectors $\eta^{1/2} \, Q R_\eta \, \Omega$, \,  
$\eta^{1/2} \, (L_\eta Q)^* \, \Omega$, \, $(1 - R_\eta) \, \Omega$ and
$(1 - L_\eta)^* \, \Omega$ tend to $0$ as $\eta \rightarrow 0$. 
Thus by taking matrix elements of relation (\ref{2.21}) in the state 
$\Omega$ and applying to the left hand side of the resulting 
equation the Cauchy--Schwarz
inequality one arrives at the statement. \hfill $\Box$\\[-2mm]

We can now establish the following result about the 
asymptotic properties of super--regular states. 

\begin{Proposition}
Let $\alpha_t, \, t\in \RR$, be asymptotically abelian. 
If $\omega$ is a super--regular state on $\cal F$, there holds 
(point-wise on $\cal A$)
\begin{equation}
\lim_{T \rightarrow \infty} 
\mbox{ $T^{-1}$} \! \int_0^T \! dt \, \omega \alpha_t \! \cdot \! 
\delta = 0. \label{2.22} \end{equation}  
In particular, all limit points 
of the net of states 
$\{ \mbox{\small $T^{-1}$} \! \int_0^T \! dt \, \omega  
\alpha_t \}_{\, T > 0}$ for $T \rightarrow \infty$ 
are supersymmetric.
\end{Proposition}
{\em Proof:\/} Because of Lemma 2.4 we have  
$ \mbox{\small $T^{-1}$} \! \int_0^T \! dt \, \omega \alpha_t (F_-)
\rightarrow 0$ for  $F_- \in {\cal F}_-$ and $T \rightarrow \infty$
and since $\delta$ is an odd derivation it follows that (\ref{2.22}) 
holds on ${\cal A}_+$. If $F_- \in {\cal A}_-$, we put
$F_{-_T} \doteq  \mbox{$T^{-1}$} \int_0^T \! dt \, \alpha_t (F_{-})$
and obtain with the help of the preceding lemma the estimate 
\begin{eqnarray}  
\lefteqn{|\mbox{ $T^{-1}$} \! \int_0^T \! dt \, \omega \alpha_t \! \cdot \!
\delta (F_-)| =  | \omega \! \cdot \! \delta (F_{-_T}) | } \nonumber \\
& & \leq  c_\varepsilon \, 
    ( || \pi (F_{-_T}) \Omega || + || \pi ({F_{-_T}}^*) \Omega||) 
+ \varepsilon \, ( || \delta (F_{-_T})|| + ||\overline{\delta}
(F_{-_T})||) \nonumber \\
& & \leq  2 c_\varepsilon \, 
    || F_{-_T} || 
+ \varepsilon \, ( || \delta (F_{-})|| + 
||\overline{\delta} (F_{-})||), \label{2.23}
\end{eqnarray} where, in the final step, 
we made use of the fact that 
$ \delta$ and $\overline{\delta}$ are linear 
and of the triangle inequality. 
Applying Lemma 2.4 another time we see that 
the first term on the right hand side of this inequality vanishes 
in the limit $T \rightarrow \infty$. Since $\varepsilon > 0$ is 
arbitrary the statement follows. \hfill $\Box$ \\[-2mm]

This result shows that if there exist in a theory super--regular 
states there exist also supersymmetric ground states which asymptotically 
approximate the regular ones. In particular, any $\alpha$--invariant
super--regular state is supersymmetric. 

In the proof of these results we made use of the assumption that $\alpha$
is asymptotically abelian, but this condition can be relaxed. 
It suffices if there is some group of automorphisms $\beta$
which acts on $\cal F$ in an asymptotically
abelian manner and commutes with $\delta$. 
If one replaces in the formulations of  
Lemma 2.4 and Proposition 2.7 the group $\alpha$ by any such $\beta$,   
the resulting statements hold as well. 

\section{Conclusions}
\setcounter{equation}{0}
In the preceding  
analysis we have seen that the implementation of 
supersymmetry transformations by Hilbert space operators can  
be accomplished only in a very special class of states. 
We want to discuss here the physical implications of 
this observation. 

In order to fix ideas let 
us assume that we are dealing with a theory with an asymptotically
abelian time evolution which commutes with the 
supersymmetry transformation $\delta$ (yet it need not 
necessarily coincide with $\alpha$).   
There then emerges the following general
picture from our results: According to  
Proposition 2.3 no thermal state is   
symmetric with respect to the action of $\delta$, for thermal states 
are faithful as a consequence of the KMS--condition. 
This holds therefore {\em a fortiori} for any mixture of 
such states. Furthermore, the action of supersymmetry cannot be implemented 
by operators on the corresponding state spaces (thermal states are 
not super--regular). This follows from Proposition 2.7, respectively 
its generalization mentioned at the end of the preceding section, 
according to which any super--regular state which is invariant under 
the time evolution is also supersymmetric. We may therefore state:\\[2mm]
{\em (a) Thermal states and their mixtures are neither supersymmetric 
nor do they admit\\ \hspace*{6mm} 
the implementation of supersymmetries by 
Hilbert space operators.\/}\\[2mm]
The fact that one cannot restore supersymmetry in thermal 
states by proceeding to suitable mixtures was termed 
spontaneous collapse in \cite{BuOj}. Our present results are slightly more 
general than those in the latter article  
since they hold without the assumption that the supersymmetry  
transformation $\delta$ is related to the generator of the time
evolution. 

It is another intriguing consequence of Proposition
2.7 that all super--regular states coincide at asymptotic times 
with stationary mixtures of supersymmetric states. 
The latter states in turn are, by Proposition 2.2 and 2.5, 
mixtures of pure (and hence in 
their respective representations unique) supersymmetric states 
which can be distinguished by central observables (macroscopic order
parameters). This general result provides evidence to the effect that
the asymptotic limits of super--regular states 
are vacuum states. 

In order to substantiate this idea let us consider 
the pertinent examples of supersymmetry in particle physics 
\cite{Lo}. There the generator $\delta_0$ of the 
time evolution can be expressed, in any given Lorentz system, 
by two odd derivations $\delta_1, \delta_2$ which commute with 
space and time translations, 
\begin{equation}
\delta_0 = \mbox{\large $\frac{1}{4}$} (\delta_1 \cdot
\overline{\delta_1} + \overline{\delta_1} \cdot \delta_1)  + 
\mbox{\large $\frac{1}{4}$} (\delta_2 \cdot
\overline{\delta_2} + \overline{\delta_2}  \cdot \delta_2).
\end{equation}
If the spatial translations act on the field algebra in an
asymptotically abelian manner (which is the case if this algebra
is generated by local fields), it follows from Proposition 2.7,
respectively its generalization, that all states which are
super--regular with respect to $\delta_1$ and $\delta_2$ 
coincide in asymptotic spacelike directions with supersymmetric
states. The latter states are, by Proposition 2.2, ground states 
for the time evolution, and this holds in all Lorentz frames
\cite{BuOj}. Hence these states are relativistic vacuum states 
which can be decomposed into pure vacuum states. (The 
requirement of asymptotic abelianess of the time evolution is not
needed here \cite[Ch.\ 2.4]{Sa}.) 
We can summarize these results as follows:\\[2mm]
{\em (b) States admitting the implementation of supersymmetry agree  
in spacelike asymp\-\hspace*{7mm}totic regions 
with (mixtures of) pure supersymmetric vacuum states.\/}\\[2mm]
Hence in a supersymmetric theory 
all super--regular states are excitations of  
vacuum states and therefore describe 
only elementary systems. More complex systems do not
admit an action of supersymmetry. 

We are led by these results to the conclusion that 
supersymmetry is extremely vulnerable to thermal effects and 
there is no way of restoring the broken symmetry by 
physical operations on the states. In contrast, 
such a restoration can in general be accomplished quite easily
in the case of broken bosonic symmetries: given a non--isotropic 
system such as a ferromagnet, say, one can prepare a corresponding 
rotational invariant (mixed) state by rotating the probe. 
As we have seen, 
there is no corresponding symmetry enhancing operation in the case 
of supersymmetry.  

In view of these facts one may wonder how supersymmetry 
manifests itself in complex physical systems, 
such as the presumed early supersymmetric stages of the 
universe, where matter has been in a hot thermal imbroglio.
It may well be that the presence of supersymmetry at the 
microscopic level of fields has no clearly visible 
consequences for such states. 
{}From the theoretical viewpoint this vulnerability of 
supersymmetry may be a virtue, however. First, it could  explain
why it is so difficult to establish this symmetry experimentally, 
should it be present in nature. Second, it might be used to distinguish 
in the theoretical setting preferred states by imposing supersymmetry
as a selection criterion. 

Thinking for example of quantum field theory
on curved spacetime manifolds which do not admit a global time evolution
(future directed Killing vector field), the notion of vacuum state becomes 
meaningless. But there might still exist in such theories 
some distinguished odd derivation. One could then 
characterize the preferred 
states and their corresponding folia by demanding that 
they be symmetric with respect to its action. 
The results of the preceding analysis would justify the view 
that such states describe the most elementary systems of the theory.  

This idea suggests the following mathematical question 
whose solution is known in the case of even derivations
\cite[Prop.\ 3.2.18]{Sa}: Under which conditions do there exist for an  
odd derivation on some graded C$^*$--algebra states which are 
annihilated by it? We hope that the present results will stimulate  
some interest in this problem.  

\end{document}